\documentclass[aps,prl,preprint,superscriptaddress]{revtex4}

\usepackage{graphicx}
\begin{document}

%\preprint{}

\title{``Super weakly" coupled superconductivity in ultrathin superconductor-normal-metal bilayers}

\author{Zhenyi Long, M. D. Stewart, Jr.}

\author{James M. Valles, Jr.}
 \email{valles@physics.brown.edu}

\affiliation{Department of Physics, Brown University, Providence, RI 02912}

\date{\today}

\begin{abstract}
Tunneling measurements of the temperature dependence of the electronic density of states (DOS) of ultrathin bilayers of Pb and Ag reveal that their superconducting energy gap, $\Delta (T)$, evolves similarly to BCS predictions despite the presence of a large anomalous DOS at the Fermi energy that persists as $T\rightarrow 0$.  The gap ratio, ${2\Delta (0)}\over {k_BT_c}$, systematically decreases more than 20\% below the BCS universal value of 3.52 in bilayers with the lowest superconducting transition temperatures, $T_c$; a behavior we deem super weakly coupled.  A semi-quantitative model suggests that the reduced gap ratios result from a systematic depletion of the DOS available for pairing that occurs with the growth of the anomalous DOS at the Fermi energy.
 	
\end{abstract}

%\keywords{}

\maketitle

The BCS theory of superconductivity predicts that the gap ratio, $\eta=2\Delta_0/k_BT_c$, assumes a universal value of 3.52 \cite{BCS}.  Here, $T_c$ and $\Delta_0$ are the superconducting transition temperature and zero temperature energy gap, respectively. Experiments on a large number of elemental superconductors show $\eta$ to range from a little below 3.5 to as high as 5 \cite{Carbotte}, in rough accord with this prediction.  Extensions of the BCS theory into the strong electron phonon coupling regime \cite{Parks} reveal that $\eta$ reflects the strength of the microscopic coupling constant and show that stronger coupling correlates with larger $\eta$ \cite{Carbotte}. This correlation is often used to classify novel superconductors as weakly or strongly coupled even when the microscopic electron pairing interactions responsible for their superconductivity are unknown \cite{TaguchiPRL2005}. Here, we present a superconducting system, ultrathin bilayers of superconductor (S) and normal metal (N), in which $\eta$ can be driven well below 3.5. We deem this surprising behavior ``super weakly" coupled superconductivity.     

Ultrathin SN bilayers, are expected to behave similarly to pure superconducting layers according to quasi-classical proximity effect theories\cite{usadel}.  Their gap ratio should conform to $\eta >$ 3.5 and their Density Of States (DOS) should exhibit an energy gap devoid of states, centered on the Fermi energy, $E_F$, provided they are thinner than the superconducting coherence length, $\xi_0$.  The presence of the N layer simply decreases the bilayer's $T_c$ \cite{Cooper}.  Tunneling experiments, however, indicate that bilayers with ultrathin S layers are not conventional. They possess an anomalous DOS within the energy gap region which becomes more pronounced with increasing normal layer thickness, $d_N$ \cite{Long,Gupta,Escoffier}. In order to fit the whole DOS it is necessary to presume the coexistence of two distinct populations of quasiparticles, one giving rise to a BCS-like DOS and the second contributing the subgap states \cite{Long}.  

Here we show that the temperature dependence of the average energy gap, $\Delta(T)$, closely resembles the BCS form indicating that the growth of the superconducting condensate is conventional and uniform through the bilayers.   In the bilayers with the lowest $T_c$'s, however, $\Delta(T)$ asymptotes to an unexpectedly low value so that $\eta<$ 3.52.  Concomitantly, the anomalous DOS at the Fermi energy appears to persist in the $T\rightarrow$ 0 limit and is largest in bilayers with the lowest $\eta$.  Present theories indicate that these states correspond to quasiparticles that are uncoupled from the superconducting phase\cite{Melsen,Schomerus,Belzig}.  By assuming that their decoupling reduces the DOS available for pairing it is possible to account for the anomalously low values of $\eta$ using simple formulae for $T_c$ and $\Delta$ and thus, the super weakly coupled behavior.      

The Pb/Ag bilayer films employed here were quench condensed from vapor onto $T$ = 8 K glass substrates in the UHV environment of a dilution refrigerator based cryostat \cite{Merchant, Kouh, LongJLTP}. Prior to cryostat mounting a substrate was cleaned, fire polished and patterned with Au/Ge contacts and Al (Alloy 2024) strips that served as the tunnel junction counterelectrodes. The quench condensed bilayers were formed by depositing an electrically discontinuous Pb film first and then a series of Ag layers. The mass per unit area of each deposition was measured with a quartz crystal microbalance and converted to a thickness using the bulk density of each element. After each Ag deposition, the tunnel junction conductance was measured as a function of temperature using standard 4-terminal, low frequency AC techniques. In this way, a series of Ag/Pb bilayers with fixed Pb thickness $d_{Pb}$ and increasing Ag thickness $d_{Ag}$ were fabricated and measured {\it{in situ}} with the same tunnel junction barrier, without breaking vacuum or warming. This paper presents data on a series of 5 bilayers with $d_{Pb}$ = 4.0 nm, 6.7 nm $<d_{Ag}<$ 19.3 nm, and 0.72 K $< T_c < $ 2.55 K. This series exhibits characteristics consistent with other series' of bilayers with 1.4 nm $< d_{Pb} <$ 4.0 nm.  

The normalized tunneling DOS of a bilayer, $\widetilde{N}_S(E)$, is related to the normalized conductance of the tunnel junctions through \cite{Tinkham}: $$G_j(V,T)=\int^{+\infty}_{-\infty}\widetilde{N}_S(E)\frac{\partial{f(E+eV)}}{\partial{(eV)}}dE$$
where $E$ is the energy measured relative to the Fermi energy, $E_F$, $\widetilde{N}_S(E)=N_S(E)/N_N(0)$ is the superconducting DOS normalized by the normal state DOS at $E_F$. $\partial{f(E+eV)}/\partial{(eV)}$ is the derivative of Fermi function and $V$ is the voltage across the junction. 

\begin{figure}
 \includegraphics{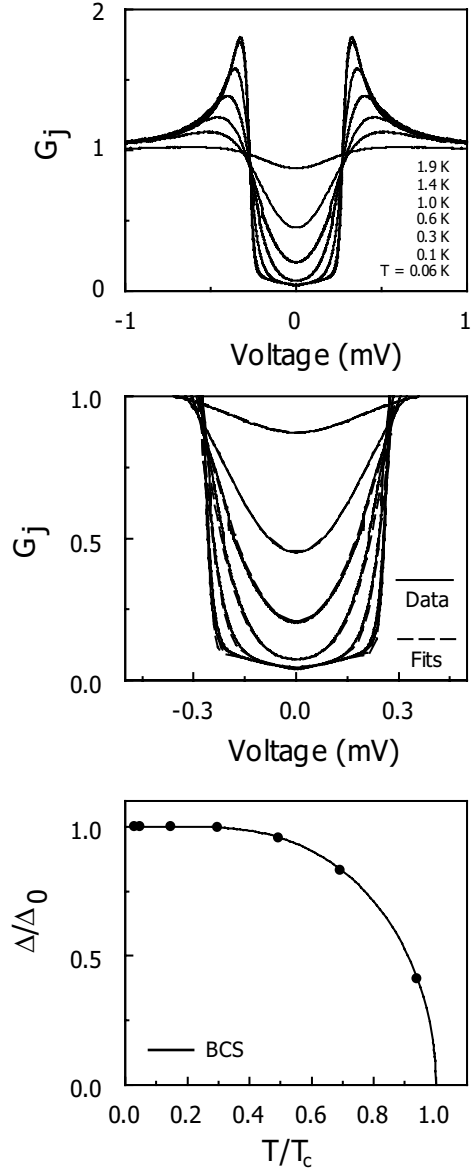}
 \caption{(a) $G_{j}$ vs. V of a bilayer film ($d_{Pb}/d_{Ag}$ = 4.0 nm/9.1 nm) at T = 0.06, 0.1, 0.3, 0.6, 1.0, 1.4, 1.9 K. (b) Data and fits in the subgap region. (c) The inferred normalized energy gap $\Delta/\Delta_{0}$ vs. $T/T_c$.}
\end{figure}

The temperature dependence of $G_j$, shown for a $T_c =$ 2.0 K bilayer in Figs. 1a and 1b, primarily follows BCS predictions with one very significant deviation. In accord with BCS, symmetric peaks grow and a depression between the peaks deepens as $T$ decreases below $T_c$. Most importantly, a well defined gap edge at $\approx \pm$0.29 meV develops at the lowest temperatures. Contrary to BCS, the depression does not drop to zero within the gapedges \cite{Tinkham}. Rather, it develops into a symmetric, linear cusp with a finite intercept (see Fig. 1b).  Experiments on other bilayers and trilayers assure us that this subgap conductance reflects an intrinsic feature of the bilayer DOS and not junction leakage \cite{Long}. Moreover, investigations by other groups on other mesoscopic hybrid structures such as thin Au/Nb bilayers \cite{Gupta} and inhomogeneous TiN films \cite{Escoffier} reveal a similarly anomalous subgap DOS.   

The behavior above suggests that the finite DOS at $E_F$ persists to arbitrarily low temperatures.  To check this further, we measured the  zero voltage bias conductance, $G_j(0,T)$, as a function of temperature.  $G_j(0,T)$ is proportional to the DOS within $\sim$4 $kT$ of $E_F$. Data for five of the bilayers with 0.72 K $< T_c <$ 2.55 K, are compared to the BCS dependence in Fig. 2a. With decreasing $T$, $G_j(0,T)$ shows an abrupt slope change at $T_c$ signalling the opening of the energy gap and subsequently, decreases monotonically in agreement with BCS.  At the lowest temperatures, however, $G_j(0,T)$ is higher than the BCS prediction and decreases nearly linearly with temperature rather than exponentially.  In fact, the $G_j(0,T)$ appear to asymptotically approach finite values. The limiting values, $G_j(0,0)$, are larger for the lower $T_c$ bilayers.      

\begin{figure}
\includegraphics{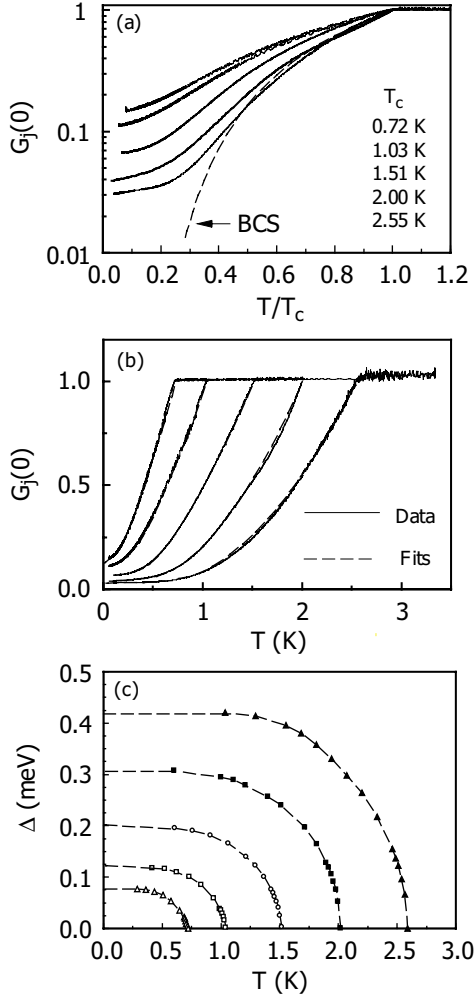}%
\caption{(a) Semi-logarithmic plot of $G_{j}(0)$ vs. $T/T_c$ for Pb/Ag bilayers with $d_{Pb}=$ 4 nm and $d_{Ag}=$ 6.7, 9.1, 12.4, 15.6, 19.3 nm. The dashed line is the mean field theoretical prediction of $G_{j}(0)$ vs. $T$. (b) Same data as in (a) with fits (dashed lines). (c)Energy gap $\Delta$ vs. $T$ obtained from the fits in b) ($d_{Pb}$ = 4.0 nm and $d_{Ag}$ = 6.7 nm (solid triangle), 9.1 nm (solid square), 12.4 nm (empty circle), 15.6 nm (empty square) and 19.3 nm (empty triangle)).}
\end{figure}

Both $G_j(V,T)$ and $G_j(0,T)$ can be fit by introducing temperature dependence into a two component form for the DOS that was introduced previously to fit tunneling data at a single low temperature $T=$60 mK $<< T_c$ \cite{Long}. This DOS consists of a broadened BCS DOS at high energies and a linear DOS with an offset at energies within the gap.  The former corresponds to BCS-like quasiparticles and the latter corresponds to quasiparticle states that are quasi-localized in the normal regions \cite{Melsen,Schomerus,Belzig}. Those contributing to the linear portion ($\alpha E$) are partially localized while those giving the offset ($\beta$) are completely localized within the N layer \cite{Long}. Explicitly,
$$\widetilde{N}_S(E)=\cases{\alpha{E}+\beta &; $E<E_{c}$\cr
 \widetilde{N}_S^{\sigma}(E) &; $E\geq{E_{c}}$}$$ 
where $E_c$ is the crossing energy separating the two DOS forms.  $\widetilde{N}_S^{\sigma}(E)$ is a superposition of BCS DOS' $N_S^{BCS}(E, \Delta)=Re(E/\sqrt{E^2-\Delta^2})$ using a log-normal distribution of $\Delta$ \cite{Long}: 
$$\widetilde{N}_S^{\sigma}(E)\!=\!\!\frac{1}{\sqrt{2\pi}(k\sigma)}\!\!\int_0^{\Delta_0}\!\!\!\widetilde{N}_{S}^{BCS}(E,\Delta)exp(-\frac{(ln(\frac{\overline{\Delta}}{\Delta}))^2}{2(k\sigma)^2})\frac{d\Delta}{\Delta}$$.
The product $k\sigma$ characterizes the width of the log-normal gap distribution and $\overline{\Delta}$ is the most probable $\Delta$. The fits to the data shown in Fig. 1b were optimized to capture the gap edges. They matched the peaks to within 10\%. The inability to capture  both features simultaneously probably results from inaccuracies in the log normal approximation for the gap distribution. $\overline{\Delta}$ was adjusted for each temperature. $\alpha$, $\beta$ and $(k\sigma)$ were presumed temperature independent and held to the values that produced the best fit at $T = $0.06 K. We are confident that this presumption is reasonable because of the physical interpretation of the parameters\cite{Long}.  It gives high quality fits to the lowest temperature data where these parameters exert their greatest influence implying that a subgap DOS persists in the $T\rightarrow 0$ limit. As shown in Fig. 1(c), $\overline\Delta(T)$ agrees well with the BCS form (the solid line) and gives an $\eta=$ 3.57, which is close to the BCS value (3.52).

Fits to the zero bias conductances for all of the bilayers, yield $\Delta(T)$ and a good estimate of the density of electronic states at $E_F$ that exists in the zero temperature limit. The fitted curves shown in Fig. 2b were calculated using: $G_j(0)=\int^{+\infty}_{-\infty}\widetilde{N}_S(E)[\partial{f(E+eV)}/\partial{(eV)}]_{V=0}dE$, presuming that $\alpha$, $\beta$ and $(k\sigma)$ are constant (see earlier discussion) and adjusting $\Delta$ at each temperature point along a curve. At low temperatures, $\Delta$ must approach a constant value, $\Delta_0$, in order for the fits to reproduce the low temperature linear dependence of $G_j(0)$. These $\Delta_0$ agree with the most probable energy gap $\overline{\Delta}$ obtained from fits to $G_j(V)$ at 60 mK to within $\sim$15\% in the worst case. The resulting $\Delta(T)$ for all of the bilayers, which are shown in Fig. 2c resemble the BCS form. Thus, the fits to $G_j(V,T)$ and $G_j(0,T)$ indicate that each bilayer has a well defined BCS-like DOS, which coexists with an anomalous finite DOS at the Fermi energy that persists in the zero temperature limit.   
\begin{figure}
 \includegraphics{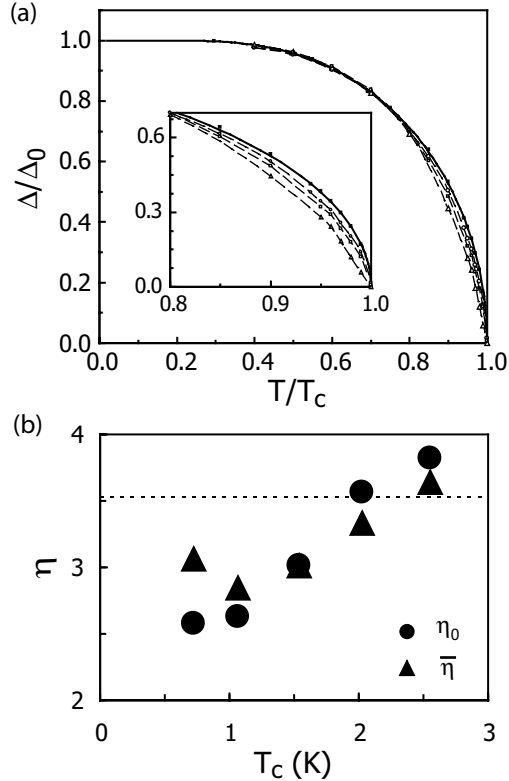}
 \caption{(a) $\Delta$ is normalized by its zero temperature value $\Delta_{0}$ plotted vs. $T$ normalized by $T_c$. The solid line is the BCS form. The inset magnifies the high $T/T_c$ range. (b) $\eta_{0}$ obtained from the fits in Fig. 2c and $\overline{\eta}$ obtained from $G_j(V)$ as a function of $T_c$. The dashed line is the BCS value for $\eta$.}
 \end{figure}

The existence of a zero temperature DOS at $E_F$ in a homogeneous s-wave superconductor without time reversal symmetry breaking perturbations is unexpected.  In principle, it could arise for bilayers outside of the Cooper limit, for which $d_N$ is thick enough that the superconducting order parameter varies perpendicular to the bilayer plane.  The exponential depression of the bilayers' $T_c$'s with normal layer thickness \cite{Kouh,Long} as well as previous work on Pb/Cu systems\cite{Ovadyahu}, however, give us confidence that the bilayers studied here are in the Cooper limit.

The ``super weak coupling" behavior becomes evident upon closer inspection of the $\Delta(T)$. First, a scaled plot, $\Delta(T)/\Delta_0$ versus $T/T_c$ reveals that the data for the three lowest $T_c$ bilayers fall below the BCS universal curve near $T_c$. These deviations, which become larger for lower $T_c$, are opposite those exhibited by strongly coupled superconductors. Second, the gap ratio, $\eta$, decreases from above 3.52 at the highest $T_c$, to well below for the lowest $T_c$ bilayers. This behavior is shown in Fig. 3b where $\eta$ has been calculated using both $\Delta_0$, $\eta_0$ and $\overline{\Delta(0)}$, $\overline{\eta}$\cite{Long}. The difference between the two reflects an uncertainty inherent in defining $\eta$ for a system with a distribution of energy gaps. Each yields a different ``averaged" value of $\eta$.   Uncertainties in $\eta$ due to $T_c$ determination are negligible for the very sharp transitions\cite{Kouh,Long}.  The relatively good agreement between the measured $\Delta(T)$ and the BCS form, which yields the extrapolated $\Delta_0$, however, leads us to believe that $\eta_0$ provides the more reliable average. Whichever the case, it is clear that $\eta$ falls well below the BCS weakly coupled value at the lowest $T_c$'s.   

Super weakly coupled superconductivity is unexpected\cite{Carbotte} although hints of it have been observed previously. Merchant and coworkers reported a steady decrease in $\eta$ to just below 3.52 in their lowest $T_c$ SN bilayers\cite{Merchant} based on extrapolations of $\Delta$ from 1.5 K to zero temperature presuming the BCS temperature dependence.  Similarly, $\eta$ systematically falls with increasing sheet resistance in amorphous Bi films near the superconductor to insulator transition \cite{Valles1994}.  In addition, $\eta<$3.52 has been reported for alloys such as $Nb_3Sn$\cite{Gregory}, but the superconducting properties of these systems have been shown to be very disorder dependent and the reduced gap ratio values were most likely due to a damaged layer at the tunnel junction interface\cite{Moore}.  

We suggest that the appearance of super weak coupling in SN bilayers is  related to the subgap DOS. Within the current model of the subgap states, $G_j(0,0)$ is proportional to the density of quasiparticles that are localized within the N regions \cite{Long,Melsen,Schomerus}. These states cannot contribute to pairing at low temperatures. They can, however, contribute to pairing at high temperatures where the divergence of the coherence length allows them to become untrapped. Thus, the effective DOS available for pairing is higher near $T_c$ than at low temperatures. This effect can reduce $\eta$ since $\Delta_0$ and $T_c$ depend on the low and high temperature DOS', respectively.  
 
\begin{figure}
\includegraphics{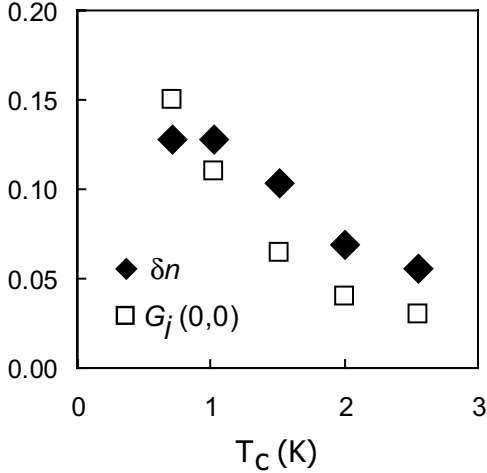}
\caption{Calculated fractional reduction in the low temperature DOS, $\delta n$, and the extrapolated zero bias conductance versus $T_c$. }
 \end{figure}

We can test the above scenario semi-quantitatively by estimating the difference in the high and low temperature DOS implied by the measured $\eta$. We use the Cooper limit relation for $T_c$ \cite{Kouh}:
$$T_c \propto exp\left(-\frac{1}{\lambda}\frac{\beta d_N+d_S}{d_S}\right)$$
where $\lambda$ is the electron-phonon coupling constant and $\beta$ is a parameter that depends on the Fermi velocities of the N and S layers and the interlayer coupling \cite{Cooper}. We take $\Delta$ to have a similar form, but with a coupling constant, $\lambda_\Delta$, that depends on $d_N$. Since the $\lambda$'s are proportional to the DOS available for pairing, the fractional difference in the high and low temperature DOS' is given by $\delta n = 1-\lambda_\Delta/\lambda$. Thus, $\eta$ and $\delta n$ are related by:
$$\delta n = 1-\left(1-\frac{d_S\lambda}{\beta d_N+d_S}log(\eta/\eta_{bulk})\right)^{-1}$$
where $\eta_{bulk}$ is the gap ratio for which $\lambda_T =\lambda_\Delta$. $\delta n$ is compared with $G_j(0,0)$ in Fig. 4. It was calculated using $\eta_0$ from Fig. 3c , the bulk Pb values $\eta_{bulk} =$ 4.38 and $\lambda =$ 0.57, and $\beta/\lambda = $ 0.4 obtained from a fit to $T_c(d_N) $\cite{Long}. The growth in the DOS at $E_F$ matches well with the predicted reduction in the density of states available for pairing and thus, supports the scenario that quasiparticle trapping leads to super weakly coupled superconductivity in these bilayers.         

The data on ultrathin Pb-Ag bilayers presented here reveal a superconducting ground state with $\eta$ decreasing below the BCS universal value as a finite DOS at the Fermi energy grows.  Our analysis suggests that these features are related and provides further support for the picture that a growing fraction of quasiparticles decouple from the superconducting phase with decreasing bilayer $T_c$ \cite{Long}.  Interestingly, $\eta$'s well below 3.5 and states within the superconducting gap \cite{Valles1994,Hsu1995} appear with decreasing $T_c$ in homogeneous thin films near the Superconductor to Insulator transition.  The spontaneous formation of non-superconducting regions embedded in superconducting regions \cite{Kowal, MasonPRB, Chervenak} may lead 
to quasiparticle localization and states at the Fermi energy in these pure, but highly disordered systems.  Taken altogether, experiments suggest that super weakly coupled behavior and quasiparticle decoupling may be general features associated with the destruction of these superconducting states.  

\begin{acknowledgments}
We acknowledge helpful conversations with R. C. Dynes and D. Feldman and the support of NSF-DMR0203608. 
\end{acknowledgments}

%\bibliography{GjT}

\end{document}